# Accurate and efficient Simulation of very high-dimensional Neural Mass Models with distributed-delay Connectome Tensors

Anisleidy González Mitjans [1,2,3], Deirel Paz Linares [1,2,4], Ariosky Areces Gonzalez [1,2,5], Min Li [1,2], Ying Wang [1,2], María L. Bringas-Vega [1,2] and Pedro A. Valdés-Sosa [1,2,4]

1. The Clinical Hospital of Chengdu Brain Science Institute, MOE Key Lab for Neuroinformation, University of Electronic Science and Technology of China, Chengdu, Sichuan, China, No. 2006, Xiyuan Ave., West Hi-Tech Zone, Chengdu,611731, China
2. School of Life Science and Technology, Center for Information in Medicine, University of Electronic Science and Technology of China, Chengdu, China
3. Department of Mathematics, University of Havana, Havana, Cuba
4. Department of Neuroinformatic, Cuban Neuroscience Center, Havana, Cuba
5. Department of Informatics, University of Pinar del Río, Pinar del Río, Cuba

*Abstract:*

This paper introduces methods and a novel toolbox that efficiently integrates any high-dimensional Neural Mass Models (NMMs) specified by two essential components. The first is the set of nonlinear Random Differential Equations of the dynamics of each neural mass. The second is the highly sparse three-dimensional Connectome Tensor (CT) that encodes the strength of the connections and the delays of information transfer along the axons of each connection. To date, simplistic assumptions prevail about delays in the CT, often assumed to be Dirac-delta functions. In reality, delays are distributed due to heterogeneous conduction velocities of the axons connecting neural masses. These distributed-delay CTs are challenging to model. Our approach implements these models by leveraging several innovations. Semi-analytical integration of the RDE is done with the Local Linearization scheme for each neural mass model, which is the only scheme guaranteeing dynamical fidelity to the original continuous-time nonlinear dynamic. This semi-analytic LL integration is highly computationally efficient. In addition to this, a tensor representation of the CT facilitates parallel computation. It also seamlessly allows modeling distributed delays CT with any level of complexity or realism, as shown by the Moore-Penrose diagram of the algorithm. This ease of implementation includes models with distributed-delay CTs. We achieve high computational efficiency by using a tensor representation of the model that leverages semi-analytic expressions to integrate the Random Differential Equations (RDEs) underlying the NMM. We discretized the state equation with Local Linearization via an algebraic formulation. This approach increases numerical integration speed and efficiency, a crucial aspect of large-scale NMM simulations. To illustrate the usefulness of the toolbox, we simulate both a single Zetterberg-Jansen-Rit (ZJR) cortical column and an interconnected population of such columns. These examples illustrate the consequence of modifying the CT in these models, especially by introducing distributed delays. We provide an open-source Matlab live script for the toolbox.

*Keywords:* Neural Mass Model, time-delay, Connectome Tensor, brain dynamics, Local Linearization Method.

## 1 Introduction

Neural Mass Models (NMM) are essential for exploring mesoscopic nonlinear brain dynamics. Obtained as mean-field approximations of ensembles of neurons, they retain sufficient simplicity to allow reasonably efficient implementation and analysis. Close cousins of these models are Neural Field models, which are not dealt with in this work but can be approximated by a sufficiently high dimensional NMM. For a review of both types of models, see (1).

NMMs serve a different purpose than models targeting the microscopic level. Programs such as GENESIS (2,3) and NEURON (4,5) allow in-silico modeling realistic neuronal aggregates. An outstanding example of this line of work is the Blue Brain (BB) (6–8) project, which aims to construct a biophysically detailed simulation of a cortical column with the most exquisite anatomical definition.

This microscopic level of detail needs to be circumvented in particular circumstances, such as comparing brain models with mesoscopic neuroimaging data. The reasoning here is that a coarser level of modeling retains the essential emergent properties of neural aggregates but allows more straightforward calculations, especially for solving inverse problems. There are excellent implementations of NMMs, examples being Dynamic Causal Modelling (DCM) (9,10), The Brain Dynamics Toolbox (11), The Virtual Brain (TVB) (12–15), among others.

This paper introduces methods and a toolbox for efficiently integrating any high-dimensional Neural Mass Model (NMM). To understand our paper's motivation, consider specific modeling and algorithmic issues that currently limit NMM capabilities. Two essential components specify these NMM.

- The dynamics of each Neural Mass: The first component is the set of nonlinear Deterministic, Stochastic, or Random Differential Equations (RDE) that specify the dynamics of each neural mass (or node).
- The second component is the highly sparse three-dimensional Connectome Tensor (CT). This tensor encodes the strength and the delay of information transfer for each connection between nodes. The three modes of the CT are emitter nodes, receiver nodes, and the distribution of delays between nodes.

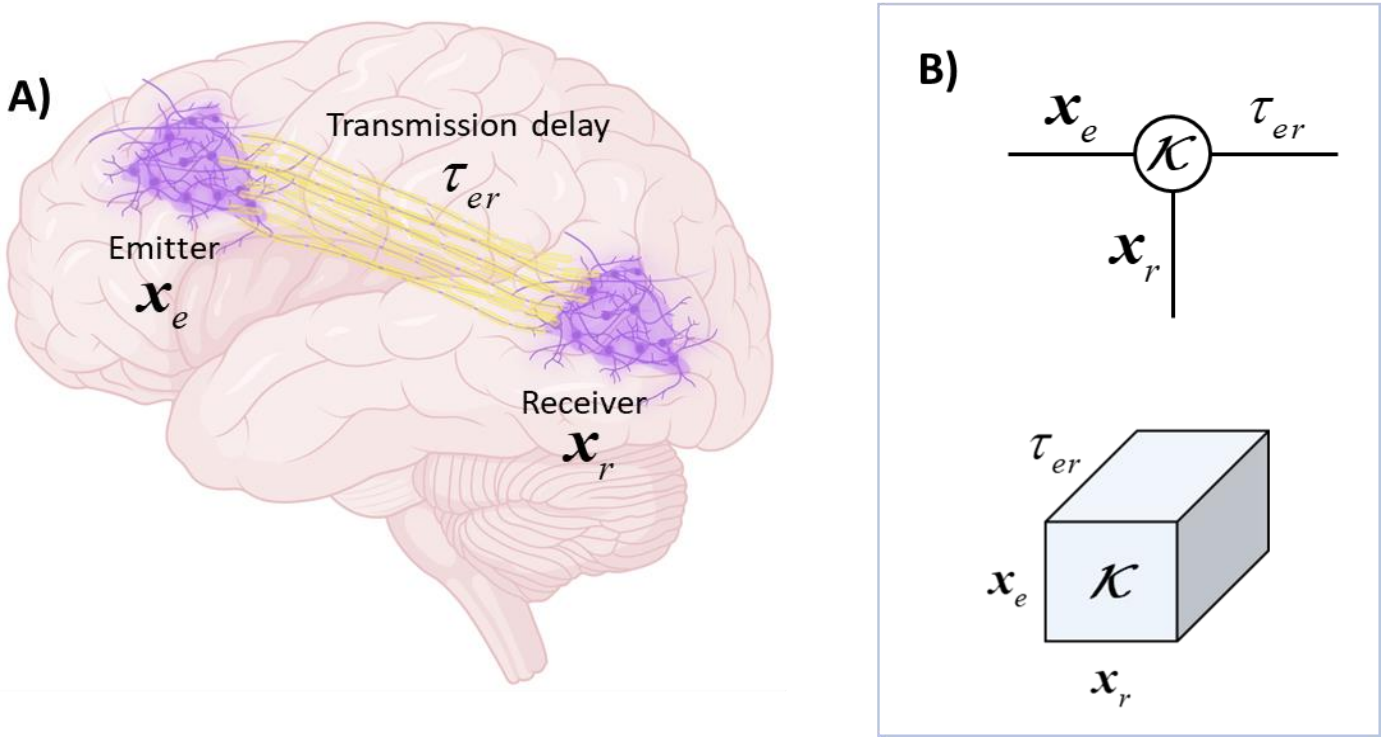

**Figure 1:** Connectome tensor of our Neural Mass Model. **A)** shows the physiological relation between two group of neural masses, where axons, emitter and receiver neuons are involved. The base of our NMM is a tensorial formulation of this phenomen, which we describe in **B)**. Figure created with BioRender.com.

Figure 1 shows the general structure of the Connectome Tensor[1]. Most work on functional connectomics collapses the CT across the temporal domain, reducing it to a connectivity matrix. Such a matrix formulation is adequate for anatomical connectivity but ignores temporal delays of signal transmission along connection pathways. The low temporal resolution of fMRI has perpetrated such an approach but is undoubtedly inadequate to disentangle causal relationships among brain regions (16).

The complexity of both node dynamics and the CT have generated conceptual and algorithmic. Here we address specific algorithmic issues that have been partially solved in successive stages. We now summarize these and explain the contributions of this paper: whose solution might enhance current NMM software capabilities. These concerns and the solutions explored in this paper are:

**Lack of Fidelity to the original dynamical system:** NMM integrates nonlinear deterministic (DE), stochastic, or random differential equations (SDE or RDE, respectively). Some current NMM implementations use standard solutions of these equations, sometimes blithely using deterministic techniques, which gives erroneous results. In the best cases, methods for SDE are employed, such as Euler-Maruyama, Heun, and 4th-order Runge-Kutta methods (17,18). These methods do not entirely preserve the original dynamics of the continuous system. Also, standard solutions become calculations frequently explosive (19–21) with probability 1. Also, for stochastic systems and small values of the integration step, the resulting Markov chain is non-stationary for nonlinear functions (22,23). A solution to these problems was introduced in (24), employing the Local Linearization Method (LLM) for the solution of SDE or RDE (20,25,26). This method guarantees both fidelity and stability to obtain a stationary Markov chain by interaction, which guarantees stationary Markov chains and non-explosive discrete-time dynamical systems, thus providing discrete-time mimicking of the original continuous-time dynamic system (27). LLM was applied for the first time to integrate NMM and parameter estimation of a neural state-space model in (28). Since then, it has been adopted in DCM (29).

---
[1] The conecept of the connectivity tensor was first proposed in (79,80).

**Excessive computational complexity:** The LL integration technique has a very high computational cost. Each integration step requires the calculation of matrix exponentials whose dimensionality is of the order of the number of neural masses. Despite the emergence of optimized methods for this calculation based on Krylov space projection (30), the numerical burden was still excessive. It limited the use of LL procedures to small-sized neural networks (20). In the paper (31), computational complexity was tackled by taking advantage of the transmission delays between neural masses. It is therefore essential to decrease the integration step of the NMMs. This increased sampling rate also increases computational cost that grows as higher-dimensional models are explored but compounded when solving inverse neuroimaging problems. Inverse problems based on NNM dynamics require iterative recalculation of state-space trajectories.

Fortunately, with a sufficiently small integration step, the input to each neural mass is already known. Thus, the solution of the RDE can be carried out for each mass in isolation. In this case, the integration for each neural mass can be solved symbolically and expressed explicitly. This tactic allows orders of magnitude acceleration of computation. The input to each neural mass can be expressed as linear combinations of time-shifted versions of the output of the other neural masses, weighted by the appropriate connectivities. This approach was used to simulate realistic, large-scale EEG and fMRI realizations in real-time (32). Here the inter-column connectivities of each Zetterberg-Jansen-Rit (ZJR) cortical column were fixed, and the long-range inter-areal connectivities were obtained from Diffusion-Weighted Imaging (DWI) (32). Despite its computational advantages, this software has not been frequently used due to its complexity and lack of detailed algorithm descriptions. It was also based on using a single conduction velocity of axonal communication between nodes.

**Difficulty in dealing with realistic profiles of axonal time delays:** When considering the delays in transmission information from one node to another, it is well known that the extreme diversity of myelination (or lack of) and axonal fiber diameters lead to a distribution of conduction velocities (and therefore delays) (33–37). The complexity of solving delay differential equations has led many teams to ignore them, assuming instantaneous axonal propagation, thus leading to systems of simultaneous differential equations. DCM finesses the issue of delays by tampering with the length of the modeled postsynaptic potential (38–40). Other groups, including TVB, pose axonal delays as Dirac-delta to facilitate solutions (41,42). This Dirac-delta delay was also used in (32), where a single conduction velocity is associated with each connection.

What we overlooked in our earlier formulation was that incorporating distributed delays could be easily achieved within the RDE-algebraic approach described in (32). We intend to remedy this omission in this paper. As mentioned previously, with a sufficiently small integration step, the inputs to each neural mass may be treated as already available, paving the way for totally general distributed transmission delays. Thus, instead of specifying connectivities between nodes as having a fixed delay, we reformulate the discrete integration scheme as a tensor operation with the three-dimensional sparse CT specifying the emitter neural masses projecting to receiver masses with unrestricted time lag distribution.

**Numerical efficiency:** As we have already stated, it is essential to decrease the integration step for the neural mass models to use the semi-analytic integration. Thus, some gains in efficiency are now offset by an increased computational cost due to more integrations. This burden grows even more as higher-dimensional models are explored. The problem is compounded when solving neuroimaging inverse problems based on NNM dynamics requiring iterative recalculation of state-space trajectories.

In this paper, we present theory and software to extend the random differential-algebraic formulation of NMM (already formulated (32)). Here, we perform in tandem, with considerable computational advantage, i) Symbolical integration of the LLM integration step for each separate neural mass and ii) Calculation of inputs to each neural mass with the highly parallelizable CT operations. We make the Matlab Live Script for the algorithm available. Additionally, we illustrate the use of the method in examples that, without pretension to neurobiological fidelity, highlight the use of the algorithm.

We show that by decreasing the integration step, inputs to each neural mass may be treated as already available, thus paving the way for totally general distributed transmission delays. We can now reformulate the discrete integration scheme as a tensor operation with the three-dimensional sparse connectivity of emitter neural masses projecting to receivers with any given lag distribution.

## 2  Model Description

Since our formulation depends on tensor operations, we briefly describe our notation summarized in Tables 1 and 2.

**Table 1:** Structure Types

| Structure | Font Description | Example |
|---|---|---|
| Tensor | Uppercase, Euclid Math One | $X \in \mathbb{R}^{I_1 \times I_2 \times \cdots \times I_N}$ |
| Matrix | Uppercase, bold, Latin | $\mathbf{X} \in \mathbb{R}^{J \times I_n}$ |
| Vector | Lowercase, bold, Latin | $\mathbf{x} \in \mathbb{R}^{I_n}$ |
| Scalar | Lowercase, Latin | $x \in \mathbb{R}$ |

**Table 2:** Operations for third-order tensors (43)

| Product | Description |
|---|---|
| Hadamard product of Tensor and matrix | $X \odot \mathbf{X} = X_{i_1 \ldots i_N} \mathbf{X}_{j i_n}$ <br> $K_{er\tau} = D_{er\tau} \odot_{er} K_{0\,er}$ |
| n-mode (matrix) product of a Tensor and matrix | $X \times_n \mathbf{X} = \sum_{i_n=1}^{I_n} X_{i_1, i_2, \ldots, i_N} \mathbf{X}_{j_n, i_n}$ <br> $\check{x}(t) = K \times_{e\tau} X^-(t)$ |

**Table 3:** Set definitions for our Neural Mass Model

| Set | Interpretation |
|---|---|
| $M = \{1, \ldots, Nm\}$ | Set of all neural masses, where $Nm$ is the total number of neural masses |
| $E = \{1, \ldots, Ne\}$ | Set of emitter nodes, where $Ne$ is the total number of emitter nodes |
| $R = \{1, \ldots, Nr\}$ | Set of receiver nodes, where $Nr$ is the total number of receiver nodes |
| $T = \begin{cases} [0, \tau_{\max}] \\ \{1, \ldots, N\tau\} \end{cases}$ | Set of transmission delays: <br> • Continuous ($\tau_{\max}$ maximum delay of information transfer between two neural masses) <br> • Discrete (with $N\tau = \text{floor}\frac{\tau_{\max}}{\Delta t}$) |
| $U = \{1, \ldots, u, \ldots, Nu\}$ | Set of ZJR cortical columns, where $Nu$ is the total number of cortical columns |

**Table 4:** Set of variables used in our Neural Mass Model

| Variable | Interpretation |
|---|---|
| $x(t)$, $x_{obs\,t}$ | State system vector and its observations, respectively |

| $\Theta$ | Dynamical system parameters determining all neural mass characteristics |
|---|---|
| $\rho(t)$ | Dynamical noise driving the neural system |
| $\xi_t$ | Measurement noise |
| $e$, $r$ | Emitter and Receivers nodes, respectively |
| $\tau_{max}$ | Maximum delay of information transfer between two neural masses |
| $\tau_{er}$ | A time delay between an emitter and a receiver node |
| $d_{er}$ | Length of the fiber system connecting an emitter and a receiver node |
| $v$ | A specific axon propagation velocity |
| $\lambda$ | Integration constant to ensure that $p_2(u;d_{er})$ is a density |
| $p(\tau_{er};d_{er})$ | Probability density of transmission delays (Examples explored in this work $p_0(\tau_{er};d_{er}) = \delta(0)$, $p_1(\tau_{er};d_{er}) = \delta\left(\tau_{er} - \frac{d_{er}}{v}\right)$ and $p_2(\tau_{er};d_{er}) = \lambda d_{er} \tau_{er}^{n-1} e^{-\alpha \tau_{er}}$ ) |
| $D$ | Delay tensor which maps emitter node $x_e$ to receiver nodes $x_r$, and a possible delay $\tau_{er}$ onto the probability density of transmission delays $p(\tau_{er};d_{er})$ in continuous or discrete time (Examples explored in this work $D_0$, $D_1$ and $D_2$) |
| $K_0$ | Connectivity matrix that does not explicitly model neural transmission delays |
| $K_{er\tau}$, $\mathbf{K}_{er\tau}$ | Connectome Tensor (continuous and discrete-time versions) |
| $x^-$ | Delayed system vector |
| $X^-(t)$ | Matrix of delayed state vectors for each neural mass |
| $\breve{x}(t)$ | Result of the past activity of all other neural masses |
| $\Delta t$ | Integration step |
| $y(t)$, $z(t)$ | Components of the state vector |
| $\mu + \sigma \dot{w}(t)$ | Noise process, |
| $s(\gamma)$ | The nonlinear sigmoid function that converts postsynaptic potentials to spiking rate |

**Table 5:** Dynamical system parameters determining all neural mass characteristics ( $\Theta$ )

| Parameter | Physiological interpretation | Pyramidal | Inhibitory | Stellate |
|---|---|---|---|---|
| $\mu$, $\sigma$ | Define the dynamical noise driving the neural system | $N(\mu_{Pyr}, \sigma_{Pyr})$ $\mu_{Pyr} = 0$, $\sigma_{Pyr} = 0$ | $N(\mu_{Inh}, \sigma_{Inh})$ $\mu_{Inh} = 0$, $\sigma_{Inh} = 0$ | $N(\mu_{Ste}, \sigma_{Ste})$ $\mu_{Ste} = 30$, $\sigma_{Ste} = 0$ $\mu_{Ste} = 30$, $\sigma_{Ste} = 1$ $\mu_{Ste} = 3$, $\sigma_{Ste} = 1$ |
| $h$ | The maximum amplitude of the PSP | $h_{Pyr} = 3.25$ mV | $h_{Inh} = 22$ mV | $h_{Ste} = 3.25$ mV |
| $\beta$ | The constant lumping together characteristic delays of the synaptic transmission | $\beta_{Pyr} = 100\, s^{-1}$ | $\beta_{Inh} = 50\, s^{-1}$ | $\beta_{Ste} = 100\, s^{-1}$ |

**Table 6:** Set of parameters to configure the Zetterberg-Jansen-Rit (ZJR) NMM

| Parameter | Physiological interpretation | Value |
|---|---|---|
| $\alpha$, $n$ | Parameters of axonal velocity distribution proposed by Nunez based on corticocortical axon diameters estimation | $\alpha = 0.6$, $n = 4.5$ |
| $\tau_p$, $\tau_i$, $\tau_s$ | Intrinsic lump parameters for excitatory and inhibitory neural masses | $\tau_p = \tau_s = 10$ ms, $\tau_i = 50$ ms |
| $e_0$, $\gamma_0$, $a$ | Parameters of the nonlinear sigmoid function | $e_0 = 0.5$ ms, $v_0 = 6$ mV, $a = 0.56$ mV$^{-1}$ |

## 2.1 General model formulation

Our general neural mass model is expressed in the state-space form:

State equation:

$$\dot{x}(t) = f(x(t), \Theta, K_0) + \rho(t) \tag{1}$$

Observation equation:

$$x_{obs\,t} = g(x(t)) + \xi_t \tag{2}$$

The state vector of all the neural masses is $x(t)$ which determines the observations $x(t) = x_{obs\,t}$ at discrete time instants with $\xi_t$ being the measurement noise. The parameters $\Theta$ determine the properties of the neural masses and the dynamical noise driving the system $\rho(t)$. Some definitions:

- We partition the state vector variables for each neural mass as $x_{(Nm \times N\tau)}(t) = \left[ x_1(t)^T, \cdots, x_j(t)^T, \cdots, x_{Nm}(t)^T \right]^T$
- Let $e \in E = \{1, \ldots, Ne\}$ denote the set of emitter nodes, $r \in R = \{1, \ldots, Nr\}$ the set of receiver nodes.
- The set of all neural masses is $M = E \cup R = \{1, \ldots, Nm\}$.

We can now define the connectivity matrix $K_{0(Nm \times Nm)}$ as the function $K_0 : E \times R \to \square$. It encodes the synaptic strength of the connections between the emitter and transmitter node. This matrix is the usual connectivity matrix of many models (17,31,44–49) that do not explicitly model neural transmission delays.

The primary purpose of this paper is to incorporate transmission delays into our models. Towards this end, we define the interval $\tau \in T = [0, \tau_{max}]$ as the set of possible transmission delays, with $\tau_{max}$ the maximum delay of information transfer between two neural masses. We also introduce the delay tensor $D_{(Nm \times Nm \times N\tau)}$ as the function $D : E \times R \times T \to \square_+$, which maps emitter node $x_e$ to receiver nodes $x_r$, and a possible delay $\tau_{er}$ onto the probability density of transmission delays $p(\tau_{er}; d_{er})$. We can now define the continuous-time Connectome Tensor ($K_{er\tau(Nm \times Nm \times N\tau)}$) with the same dimensions as $D$ via the Hadamard product:

$$K_{er\tau} = D_{er\tau} \square_{er} K_{0\,er} = D_{er\tau} K_{0\,er} \tag{3}$$

The $D$ tensors used in this paper are:

- $D_{1(Nm \times Nm \times N\tau)}$, which takes $p_1(\tau_{er}; d_{er}) = \delta(\tau)$ a Dirac-delta centered at 0. Then $K = K_0$ and $\tau_{max} = 0$, simplifying to the usual no-delay connectivity model.

- $\boldsymbol{D}_{2(Nm \times Nm \times N\tau)}$, takes $p_2(\tau_{er}; d_{er}) = \delta\left(\tau_{er} - \frac{d_{er}}{V}\right)$. In this case, $d_{er}$ is the length of the fiber system connecting nodes $x_e$ and $x_r$ a single axon propagation speed. Examples of this approach can be found in (31,50). This approach ignores the diversity of axon diameter and degree of myelination in reality but is simple to estimate.

- A more realistic $\boldsymbol{D}_{3(Nm \times Nm \times N\tau)}$, takes $p_3(\tau_{er}; d_{er}) = \lambda\, d_{er}\, \tau_{er}^{n-1} e^{-\alpha \tau_{er}}$, where $\lambda$ is an integration constant to ensure that $p_3(u; d_{er})$ is a density, $\alpha = 0.6$, $n = 4.5$, and $d_{er}$ is defined as above. This distribution function was proposed by Nunez (51) based on an experimentally determined distribution of corticocortical axon diameters (51).

Other delay models are possible and easily incorporated into our framework.

We are now ready to state our Neural Mass Model with distributed-delay Connectome Tensor expressed as

$$\dot{\boldsymbol{x}}(t) = f(\boldsymbol{x}(t), \boldsymbol{x}^-, \Theta, \boldsymbol{K}) + \boldsymbol{\rho}(t) \tag{4}$$

where $\boldsymbol{x}^-$ is the delayed system vector and $\boldsymbol{K}$ is the CT defined above.

We can now explain the most significant simplification underlying our proposed algorithms, and it is that we can perform many computations for each neural mass separately. To proceed, we need the following definitions:

- We denote $\boldsymbol{X}^-_{(Nm \times N\tau)}(t) = \left[\boldsymbol{x}_1^-(t)^T, \cdots, \boldsymbol{x}_j^-(t)^T, \cdots, \boldsymbol{x}_{Nm}^-(t)^T\right]^T$ the $M \times T$ matrix of delayed state vectors for each neural mass where $\boldsymbol{x}_j^- = \{\boldsymbol{x}_j(t-\tau) \mid \tau \in T\}$.

- $\breve{\boldsymbol{x}}_j(t)$ is the input to a neural masses $j$ of the past activity of all other neural masses.

With the definitions in place, we can define our simplified neural mass model as

$$\begin{aligned}
&a)\ \boldsymbol{x}(t_0) = \boldsymbol{x}_0 \\
&b)\ \dot{\boldsymbol{x}}_j(t) = f_j(\boldsymbol{x}_j(t), \breve{\boldsymbol{x}}_j(t), \Theta_j) + \boldsymbol{\rho}_j(t) \\
&c)\ \breve{\boldsymbol{x}}(t) = \boldsymbol{K} \times_{e\tau} \boldsymbol{X}^-(t)
\end{aligned} \tag{5}$$

This formulation substitutes the set of random differential delay equations (4) with a set of random differential-algebraic equations. This approach generalizes the differential-algebraic interpretation from (32), introducing generalized distributed delays and making calculations more accessible and less time-consuming.

*2.1.1 Numerical Integration*

To deal with numerical computations, we must discretize the system of equations (5). We discretize the connectivity tensor by defining an integration step $\Delta t$. $T$ is now sampled at multiples of $\Delta t$. This discretization allows us to redefine $\tau \in T = \{1, \ldots, N\tau\}$ with $N\tau = \text{floor}\frac{\tau_{\max}}{\Delta t}$. We now discretize the Connectome Tensor as:

$$\boldsymbol{K}^d_{e r \tau} = \int_{\tau \Delta t}^{\tau(\Delta t+1)} u\, p_i(u; d_{er})\, du \tag{6}$$

$\boldsymbol{K}^d_{e r \tau (Nm \times Nm)}$ is the discrete-time Connectome Tensor, $p_i(u; d_{er})$ is the probability density of transmission delays for each delay model $i = 0, 1, 2$ defined above, and $[t\Delta t; t(\Delta t + 1)]$ is the integration interval.

To integrate equation (5)b) we apply the Local Linearization Method (LLM) (24,32,52,53), which overcomes the limitations of the classical methods mentioned in the introduction. This method is applicable regardless of the complexity of the neural mass in question. Our approach constructs an equivalent locally linear system for the interval $[t \cdot \Delta t; t \cdot (\Delta t + 1)]$ as described in (26).

The resulting discretization $x_n = x(t_n) = x(t_0 + n \Delta t)$ of the model is given by

$$x_{n+1} = x_n + \phi(t_n, x_n, \breve{x}_n, \Theta, \rho) \tag{7}$$

The integration starts with the initial value $x(t_0) = x_0$.

We obtain the integration function $\phi$ is theoretically based on the use of the matrix exponential (54,55). However, we efficiently carry out this computation through a specific block triangular matrix combining various specifically designed submatrices. See (56) for the method description. Consequently, the discrete state-space model is simply

$$x_{n+1} = x_n + L e^{Jf \Delta t} q \tag{8}$$

where $q = [0_{1 \times (d+1)}, 1]^T$, $L = [I_d, 0_{d \times 2}]$.

Henceforth, we specialize the general model in equations (5) to the case where each neural mass follows the dynamics stated in the Zetterberg (45)-Jansen-Rit (57) (ZJR) NMM family of models. However, we note that our proposals may be used with other formulations. We note that there are two standard formulations for ZJR NMM. One uses kernels as done in (58–63). The other form uses the model's state-space form (32), which we follow in this work.

*2.2   Model for a single neural mass and algebraic integration*

For a single ZJR neural mass model formulation, equation (5)b) can be rewritten as a first-order system of two differential equations with the vector input $\breve{x}_j(t)$ and the vector output $x_j(t)$. The neural mass state vector takes the form $x_j(t) = \begin{bmatrix} y_j(t) \\ z_j(t) \end{bmatrix}$, where $y_j(t)$ and $z_j(t)$ are scalar functions. We follow the state space specification developed in (64). Therefore, following the statement of equation (5)b,c), the model for a single neural mass is formulated as

$$\begin{cases} \dot{y}_j(t) = z_j(t) \\ \dot{z}_j(t) = h_j \beta_j \left[ \mu_j + \sigma_j \dot{w}_j(t) + s_j(\breve{y}_j(t)) \right] - 2\beta_j z_j(t) - \beta_j^2 y_j(t) \end{cases} \tag{9}$$

where $y_j(t)$ is the primary variable, which represents the PSP output of a single neural mass and $\dot{z}_j(t)$ is its derivative.

The system parameter vector for a single neural mass is $\theta_{j(1 \times 4)} = [\mu_j, \sigma_j, h_j, \beta_j]$ with $j \in M$, The values of these parameters determine the system response of the neural mass (See Table 4). The system has two inputs: the noise process $\mu_j + \sigma_j \dot{w}_j(t)$ and an external input process $\breve{y}_j(t)$ (from other neural masses).

We show the block diagram of a single neural mass in Figure 1. $h_j$ is the maximum amplitude of the PSP and $\beta_j$ is the time constant of the membrane involved in the model, which depends on the rate constants of passive membrane and other spatially distributed delays in the dendritic tree (58).

$s_j(\breve{y}_j(t))$ refers to the nonlinear sigmoid (the next potential-to rate) function that determines the pulse density and transforms the average membrane potential of a neural population into an average firing rate (57,65). It is defined as

$$s_j(\gamma_j) = \frac{2e_0}{1 + e^{a(\gamma_0 - \gamma_j)}} \quad (10)$$

where $e_0$ is the maximum average firing rate of the neural populations, $\gamma_0$ is the PSP corresponding to $e_0$, and $a$ is the slop of the sigmoid. The values of these parameters depend on whether the PSP is excitatory (EPSP) or inhibitory

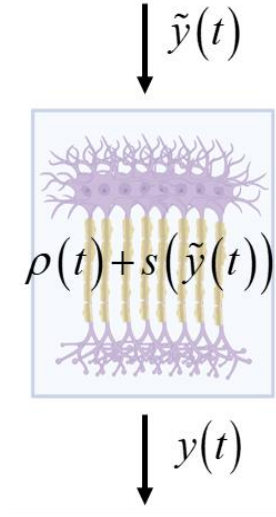

**Figure 2:** This figure describes the dynamics in a single neural mass. The random inputs to the neural mass are collected in the term $\rho_j(t) = \mu_j + \sigma_j \dot{w}_j(t)$ and $s_j(\breve{y}_j(t))$ is the usual sigmoid function that transforms the membrane potentials of a neural population into axonal output firing rates. Figure created with BioRender.com.

(IPSP) and can be found in Table 3 (66–68).

Since the system is small enough to be solved symbolically, we carried out the algebraic derivation of the LL integration step explicitly with a Live Script using the MATLAB Symbolic Toolbox. The output of these operations is shown in the expressions (12)-**Error! Reference source not found.**, where we group the random system input into the variable $\rho_j(t)$.

$$Jf_{x_j} = \begin{bmatrix} 0 & 1 \\ -\beta_j^2 & -2\beta_j \end{bmatrix} \quad (11)$$

$$Jf_{\binom{\breve{y}_j}{\rho_j}} = \begin{bmatrix} 0 & 0 \\ h_j \beta_j \dfrac{d}{d\breve{y}_j}\left(s_j(\breve{y}_j(t))\right) & h_j \beta_j \end{bmatrix} \quad (12)$$

Furthermore, the full Jacobian of the system is determined substituting the expressions (5)b), (12) and **Error! Reference source not found.** obtained by algebraic derivation as follows

$$Jf = \begin{bmatrix} Jf_{x_j}(x_{j,n}, \rho_{j,n}) & Jf_{\rho_j}(x_{j,n}, \rho_{j,n}) \dfrac{\rho_{j,n+1} - \rho_{j,n}}{\Delta t} & f(x_{j,n}, \rho_{j,n}) \\ 0 & 0 & 1 \\ 0 & 0 & 0 \end{bmatrix} \quad (13)$$

In the next step of the LLM, after applying the exponential matrix, is a regrouping of terms was made, which simplifies and facilitates the compression and manipulation of the following numerical expression

$$x_j(t + \Delta t) = (I + A_j)x_j(t) + B_j e_j(t) \quad (14)$$

where $A_j$ and $B_j$ are constant matrices with the PSP factors for an arbitrary neural mass $j \in M$, $I$ is the identity matrix, and $e_j(t)$ contains the random inputs to the system and the sigmoidal function (32).

$$A = \begin{bmatrix} \exp(-\beta_j h_j)(\beta_j \Delta t + 1) - 1 & \exp(-\beta_j \Delta t) \Delta t \\ -\beta_j^2 \exp(-\beta_j \Delta t) & \exp(-\beta_j h_j)(1 - \beta_j \Delta t) - 1 \end{bmatrix} \quad (15)$$

$$B_j = \begin{bmatrix} \dfrac{h_j \exp(-\beta_j \Delta t)(-\beta_j \Delta t + \exp(-\beta_j \Delta t) - 1)}{\beta_j} & \dfrac{h_j \exp(-\beta_j \Delta t)(\beta_j \Delta t + \exp(-\beta_j \Delta t)(\beta_j \Delta t - 2) + 2)}{\beta_j^2 \Delta t} \\ \beta_j h_j \exp(-h_j \Delta t) \Delta t & \dfrac{h_j \exp(-\beta_j \Delta t)(-\beta_j \Delta t + \exp(\beta_j \Delta t) - 1)}{\beta_j \Delta t} \end{bmatrix} \quad (16)$$

$$e_j(t) = \begin{bmatrix} \rho_{j,t+\Delta t}(t) + s_j(\breve{y}_j(t)) \\ \rho_{j,\text{diff}}(t) + \breve{y}_{j,\text{diff}}(t) \dfrac{d}{d\breve{y}_j}(s_j(\breve{y}_j(t))) \end{bmatrix} \quad (17)$$

Here we consider $\rho_{j,\text{diff}}(t) = \rho_{j,t+\Delta t}(t) - \rho_{j,t}(t)$ and $\breve{y}_{j,\text{diff}}(t) = \breve{y}_{j,t+\Delta t}(t) - \breve{y}_{j,t}(t)$. Since $\dfrac{ds_j}{d\breve{y}_j}(0) = 0$ we reduce the expression (17) to

$$e_j(t) = \begin{bmatrix} \rho_{j,t+\Delta t}(t) + s_j(0) \\ \rho_{j,\text{diff}}(t) \end{bmatrix} \quad (18)$$

Note that the single neural mass can be considered from now on as a "black box", a node of any ZJR-based neural mass model which can ignore the internal details (such as the mechanics of the numerical integration). It remains now to assemble different models with these nodes.

*2.3 Single Zetteberg-Jansen-Rit (ZJR) Cortical Column Level*

Our model framework now can describe the original Zetterberg – Jansen-Rit (ZJR) model (46,47) for a single cortical column. The ZJR column is a system composed of three neural masses (Pyramidal: $\text{Pyr}$, Inhibitory: $\text{Inh}$ and Stellate: $\text{Ste}$).

Consequently, the state vector of the model $x(t) = \begin{bmatrix} x_{\text{Pyr}}(t)^T, x_{\text{Inh}}(t)^T, x_{\text{Ste}}(t)^T \end{bmatrix}^T$ comprises the state vector of the three neural masses involved ($x_{\text{Pyr}}(t)$, $x_{\text{Inh}}(t)$ and $x_{\text{Ste}}(t)$). Each of these is a first-order system of two differential equations (9), thus resulting in six random differential equations.

We rewrite our general neural mass formulation as

$$\begin{aligned} &a) \; x(t_0) = x_0 \\ &b) \; \dot{x}_j(t) = f_j(x_j(t), \breve{x}_j(t), \Theta_j) + \rho_j(t) \quad j \in \{\text{Pyr}, \text{Inh}, \text{Ste}\} \\ &c) \; \breve{x}(t) = K \times_{e\tau} X^-(t) = K_0 \cdot \begin{bmatrix} X^-_{\text{Pyr}}(t) \\ X^-_{\text{Inh}}(t) \\ X^-_{\text{Ste}}(t) \end{bmatrix} \end{aligned} \quad (19)$$

The neural masses in a single ZJR cortical column have different properties and are characterized by the parameter matrix $\Theta_{4\times 3} = \begin{bmatrix} \theta^T_{\text{Pyr}}, \theta^T_{\text{Inh}}, \theta^T_{\text{Ste}} \end{bmatrix}$ see Figure BB, which contains the model parameters of the three neural masses. The system inputs act on the Stellate cells, and the neural activity comes out from the Pyramidal cells. See Table 4 for further description.

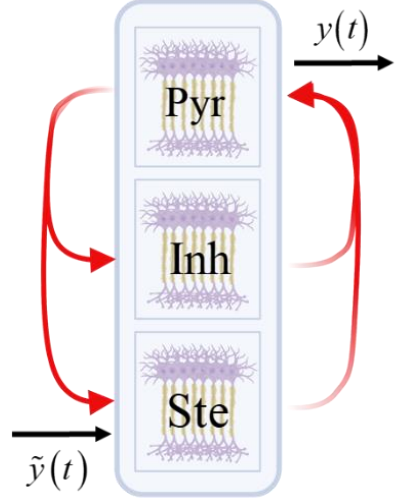

**Figure 3:** This a scheme of a single ZJR cortical column that include three neural masses (Pyramidal: $\text{Pyr}$, Inhibitory: $\text{Inh}$ and Stellate: $\text{Ste}$). The red arrows describe the intrinsic connections between neural masses, while the black arrows represent the inputs to the cortical column affecting the Pyramidal neural mass and the outputs via the Stellate neural mass.

For a single ZJR cortical column, the delay tensor $\boldsymbol{D}_0$ is modeled by $p_0(\tau_{er}; d_{er}) = \delta(0)$, that is, only with local intra-columnar connections, and we assume that the transmission time is faster than the sampling rate. Therefore, the Connectome Tensor reduces to $\boldsymbol{K} = \boldsymbol{D}_0 \square \boldsymbol{K}_0 = \boldsymbol{K}_0$. In this case, the connectivity matrix $\boldsymbol{K}_0 = \boldsymbol{C}$ contains the weights of the directed connections between the neural masses, which are well established in the literature to generate EEG-like activity (57,69). Furthermore, the synaptic connectivity between Excitatory (positive weights) and Inhibitory (negative weights) cells is shown below

$$\boldsymbol{C} = \begin{pmatrix} 0 & -33.75 & 108 \\ 33.75 & 0 & 0 \\ 135 & 0 & 0 \end{pmatrix} \quad (20)$$

Expressing the well-known ZJR cortical column with our elaborate formulation might seem a trivial exercise. However, it does prove that model includes the simpler cases. When dealing with more complex models, such as those described in the next section, the advantages of the general model might become more evident.

*2.4    Population-Level model as the result of the interactions between several ZJR-cortical columns*

Some literature explores a double-column NMM (57,70,71). We thought it instructive to test our NMM with a larger number of cortical columns (Nu=1000), a model we call "Population-Level". We now need to label each ZJR cortical column with a superscript ($u \in U$) and a subscript to any quantity that refer to the three types of neural mass in each column (i.e., $j \in \{\text{Pyr}, \text{Inh}, \text{Ste}\}$). For example, the system parameter matrix for the unit $u$ is

$$\boldsymbol{\Theta}_{3 \times 4}^u = \left[ \left( \boldsymbol{\theta}_{\text{Pyr}}^u \right)^T, \left( \boldsymbol{\theta}_{\text{Inh}}^u \right)^T, \left( \boldsymbol{\theta}_{\text{Ste}}^u \right)^T \right]^T$$

The complete system parameters are contained in the matrix $\boldsymbol{\Theta}_{3\text{Nu} \times 4} = \left[ \left( \boldsymbol{\Theta}^1 \right)^T, \ldots, \left( \boldsymbol{\theta}^u \right)^T, \ldots, \left( \boldsymbol{\theta}^{Nu} \right)^T \right]^T$.

This time, the probability density of transmission delays is $p_2(\tau_{er}; d_{er}) = \lambda d_{er} \tau_{er}^{n-1} e^{-\alpha \tau_{er}}$, which means that the delay tensor corresponds to $\boldsymbol{D}_2$.

We build the Connectivity matrix $\boldsymbol{K}_0$ manually, satisfying that

$$\boldsymbol{K}_{0\,ij} = \begin{cases} 0, & j = i \\ \boldsymbol{C}_s, & j \in \{i+1,\ldots,Ns\} \cup \{Nu - Ns +1,\ldots, Nu\} \\ \boldsymbol{C}_l, & j \in \{i+Ns+1,\ldots, Nu - Ns\} \end{cases} \quad (21)$$

where the submatrices $\boldsymbol{C}_s$ and $\boldsymbol{C}_l$ define the short-range and the long-range intercolumn connectivity, respectively. The scalar $Ns = \text{floor}\left(\frac{Nu}{6}\right)$ represents the number of sample points for the short-range connectivity and $Nu$ is the total number of cortical columns involved in the model with $i, j \in U$.

To model short-range connections we use an exponential function (31,59), defined as

$$\varphi(t) = \exp\left(\frac{-|x|}{b}\right) \quad (22)$$

where $b$ is a constant. We determine the long-range connections randomly, which could be extracted from diffusion analysis.

Here, we illustrate with an example of five ZJR cortical columns (i.e., $Nu = 5$) how the nodes are connected

Therefore, once the connectivity matrix is established, we can compute the CT as $\boldsymbol{K}_{er\tau} = \boldsymbol{D}_{er\tau} \square_{er} \boldsymbol{K}_{0\,er}$.

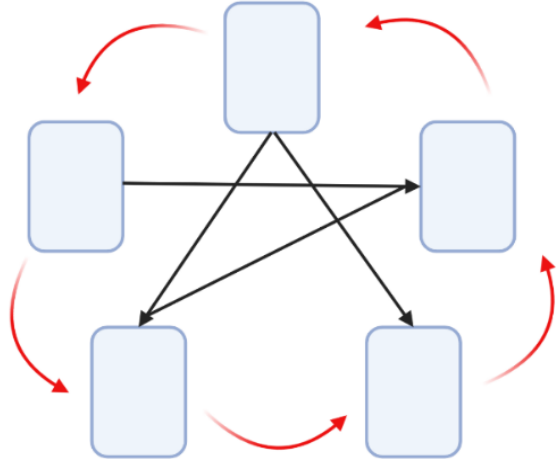

**Figure 4:** This figure shows how five ZJR cortical columns are connected, where the double arrows in red and black color symbolize the short-range and the long-range connections, respectively.

Consequently, the ADE $\breve{y}_j(t) = \boldsymbol{K} \times_{e\tau} y_j(t)$ is expanded as

$$\breve{y}_e(t) = \sum_{r=1}^{R} \boldsymbol{K}_{0\,er} y_r(t - \tau_{er}) \quad (23)$$

where $\tau_{er}$ is the time lags for two connected neural masses.

Afterward, we calculated the output of the model at the Population Level as

$$\breve{y}_e^u(t) = \sum_{r=1}^{Nm} \sum_{\tau_{er}=1}^{N\tau} \boldsymbol{K}_{er\tau_{er}} y_r^u(t - \tau_{er}) \quad (24)$$

The modeling of a single cortical column is a particular case of the Population Level, where $\boldsymbol{K}_0$ coincides with the intrinsic connections in a cortical column, i.e., $\boldsymbol{K}_0 = \boldsymbol{C}$.

## 3   Definition of Connectivity Matrix topologies $K_0$ and conduction delays for NMM simulations

We explore the effect on neural dynamics of three different types of Connectivity Matrix topologies $K_0$ and delays. For this purpose, we assume that the neural masses are distributed on a ring in two dimensions.

- For the first Connectivity Matrix topology, we specified short-range connectivity between nearest-neighbor nodes on the ring. Large-range connections were not considered, i.e., $C_l = 0$ (See Figures 5A1, A2). We call it NN network.
- We keep short-range connections to neighbors in the second case, but we define sparse large-range connectivities (See Figures 5B1, B2). We call it SW network.
- The last choice is a fully connected graph, i.e., $\forall i, j \in U$, $C_s(i,j) \neq 0$ and $C_l(i,j) \neq 0$ (See Figures 5C1, C2). We call it FC network.

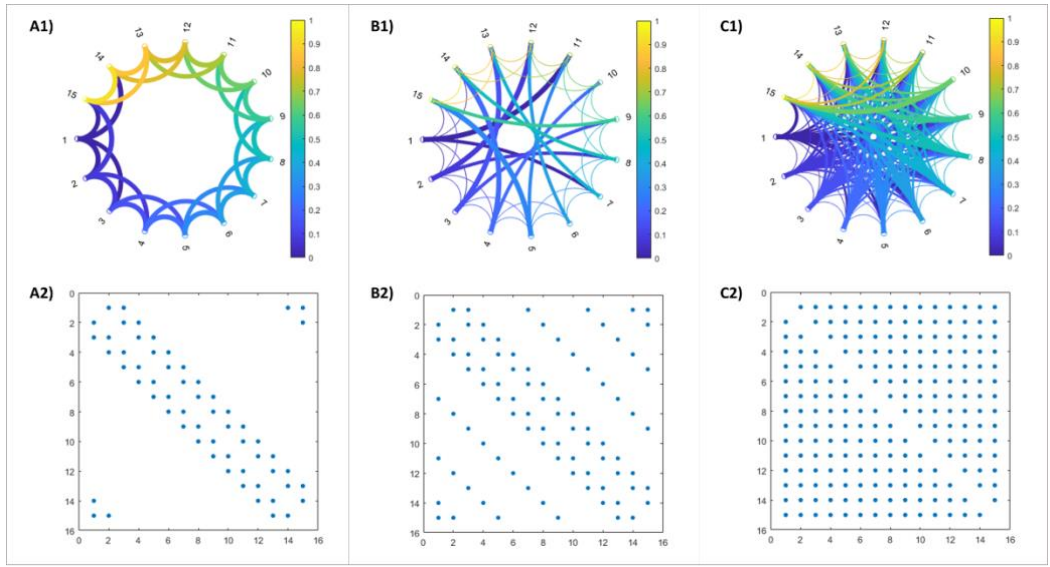

**Figure 5**: An example of the Connectivity Matrix topology for 15 ZJR cortical columns placed on a ring. The first row displays a circular graph visualization (Created with circularGraph), while the second one illustrates the sparsity pattern of the Connectivity Matrix. The first column (**A1**, **A2**) corresponds to neaerest-neibhbor connections, without large-range connections. The second column (**B1**, **B2**) Adds sparse large-range connections to the nearest-neighbor ones. The third column (**C1**, **C2**) is a full connected topology.

For all simulations, delays within a column are instantaneous. We explored the three different types of delay tensors $D$ described in section 2.1 and illustrated in Figure 6:

- $D_1$ (instantaneous delay tensor) is a Dirac-delta function centered at $0$ (Figure 6A).
- $D_2$ (synchronized delay tensor), a Dirac-delta function centered at $7.5$ (Figure 6B).
- $D_3$ (distributed delay tensor), the gamma distribution function for axonal delays proposed by Nunez (51,72), where the probability density function for corticocortical propagation peaks is about $6-9 \text{ m/s}$, with a half-width of the distribution estimated to be about $3 \text{ m/s}$ (Figure 6C).

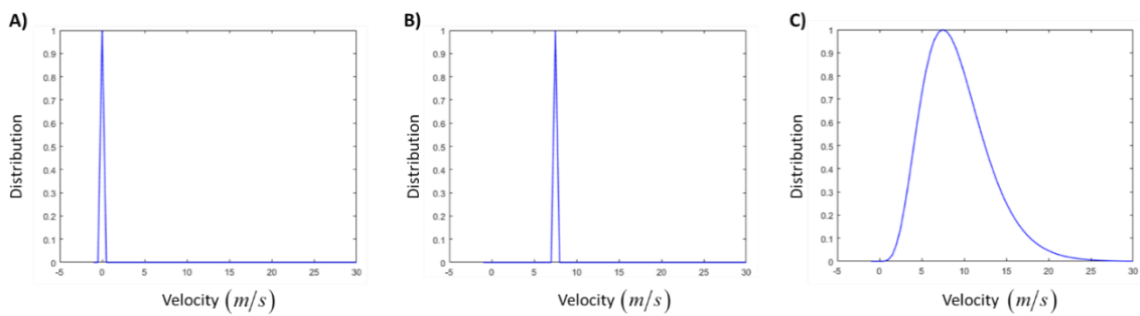

**Figure 6:** We explore three types of propagation velocity to determine the conduction delays and here we show how is the function behavior. The first way (**A**) is using $D_1$, for all connections a Dirac delta function centered in $0$, while the second one (**B**) $D_2$ is also a Dirac delta distribution for all connections but this time is centered in $7.5$. The third choice (**C**) $D_3$ is via a distribution proposed by Nunez.

## 4 Results

### 4.1 Simulation of a single ZJR cortical column

The classical ZJR NMM of a single cortical column comprises three neural masses (See Figure 7) with a delay tensor $D_1$ corresponding to zero delays. The simulation used a delta of 1 msec, and a burn-in period of 0.5 sec. After the burnout, a segment of 1 sec was retained for the analyses. In what follows, we analyze the output of the Pyr neural mass.

At this level, we tested 3 different scenarios for our mesoscale simulations. Each simulation scenario used the values for the dynamic noise, as shown in Table 5. The first (determinstic) scenario has a relatively high mean and rather small standard deviation, i.e. $\mu = 30, \sigma = 0$ (Figures 7A1, A2, and A3). Our second scenario has the same mean but a small value for the standard deviation, i.e. $\mu = 30, \sigma = 1$ (Figures 7B1, B2, and B3), and the third scenario retains the small standard deviation but also has a small value for the mean, i.e. $\mu = 3, \sigma = 1$ (Figures 7C1, C2 and C3). Figure 7 displays an example of the 3 simulation scenarios.

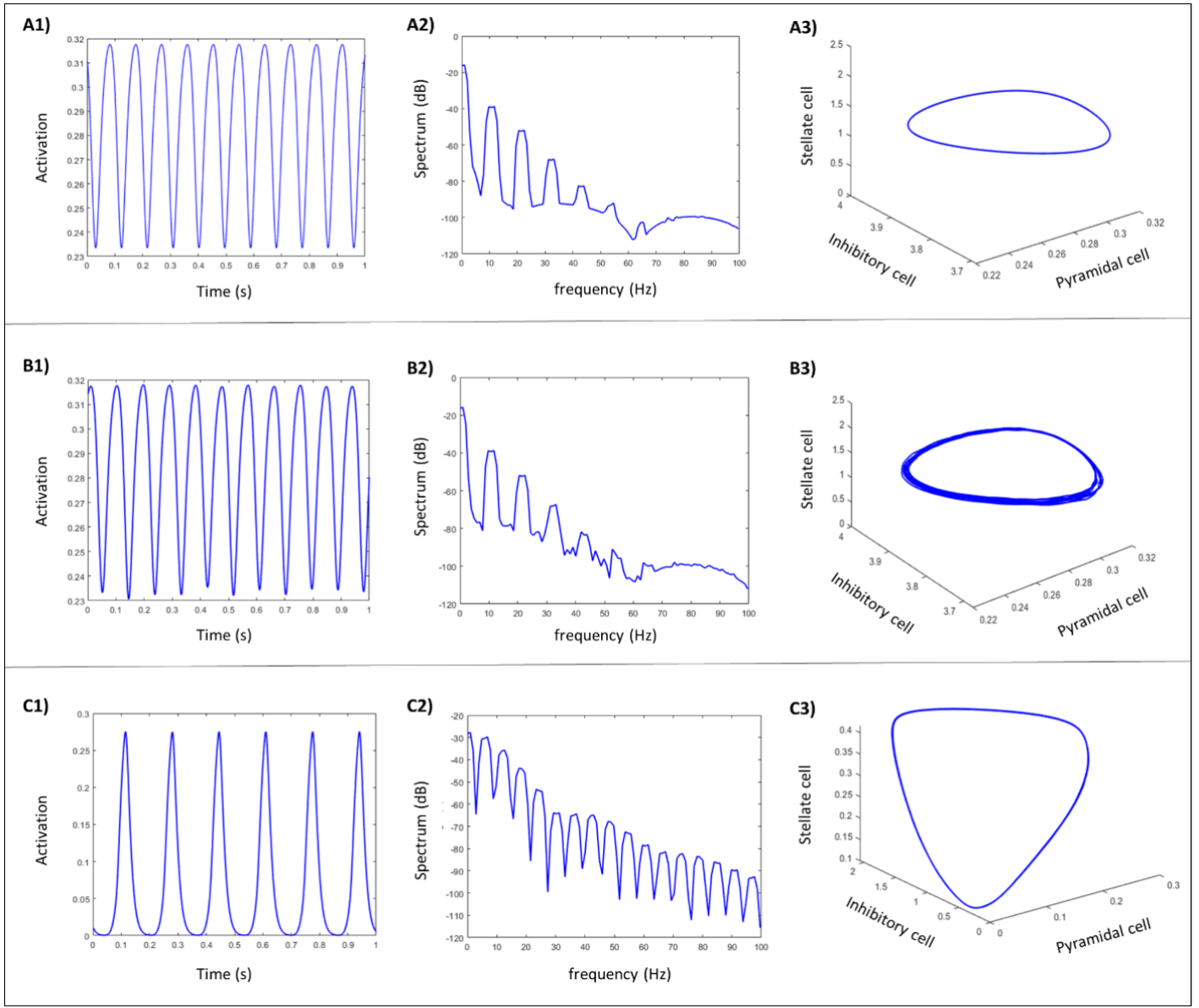

**Figure 7:** This is a three-panel figure of a single ZJR cortical column simulation for three examples of dynamic noise values. Results are shown in the left and middle columns for the model output (Pyr). The right column shows the state spce represnatation of the activity of the 3 neural masses (Pyr, Inh, Ste). Top row: simuaton without noise, i.e., $\mu = 30, \sigma = 0$, Middle row : the simulation with a small increase in the standard deviation i.e., $\mu = 30, \sigma = 1$, and Bottom row: mean and variance take small values, i.e., $\mu = 3, \sigma = 1$. **A1**, **B1** and **C1** are the times series generated by P by the ZJR NMM with parameters in Table 5 and 6, The middle column (**A2**, **B2** and **C2**) is the power spectra of the left column. The right column (**A3**, **B3** and **C3**) is the state space representation of the ZJRNMM.

Our analyses are carried out to simulate neural mass output Pyr time series and estimate the corresponding spectrum by the multi-taper method from the Chronux Matlab package (73,74). The rightmost column shows the 3D of phase plot for the output of the 3 neural masses: Pyr, Inh, Ste.

The simulated Pyr time series for all three simulation scenarios are smoothly periodic. As we can be seen, the first two simulation scenarios are pretty similar, showing only light perturbations in the spectrum and into the neural dynamics of the second simulation (See Figures 7B2 and B3). These perturbations are due to introducing a small noise level in the NMM.

We observe that for the first two scenarios (i.e., $\mu=30, \sigma=0$ and $\mu=30, \sigma=1$), the spectrum loses the harmonics for frequencies higher than $60\text{Hz}$, while the spectrum preserves more harmonics for the third scenario (i.e. $\mu=3, \sigma=1$). As was expected, the dynamics of the three simulation scenarios all converge to a limit cycle in the state space plot, regardless of the dynamical noise values explored.

### 4.2 Simulation of several ZJR cortical columns

We then tested our toolbox with an extended neural mass network (eZJR NMM). The number of cortical columns was increased to 1000 columns = 3000 neural masses. We restricted attention hereafter to the first (deterministic) simulation scenario $\mu=30, \sigma=0$. Here we explored the effect of changing two aspects of the Connectome Tensor.

1. We investigate whether the eZJR NMM is affected by the topology of the connectivity matrix topology ($\boldsymbol{K_0}$).
2. We examine the influence of different types of delay tensor distributions ($\boldsymbol{D}$) on the dynamics of the ZJR NMM.

### 4.2.1 Effects of changes in the topology of the Connectivity Matrix topology ($\boldsymbol{K_0}$) on the eZJR NMM

For these simulations, we fix the delay tensor $\boldsymbol{D}_3$ (Figure 6C). We consider three different topologies of connections.

- **The NN network:** This type of network comprises only short-range connectivity between nearest-neighbor nodes without large-range connections (Figures 5A1, A2)
- **The SW network**: A "small world" type of networking retains short-range connections to neighbors and includes sparse large-range connectivities (Figures 5B1, B2)
- **The FC network**: A fully connected graph (Figures 5C1, C2)

The analysis of the output eZJR model is similar to the one made for a single ZJR cortical column. However, instead of looking at the output of a single ZJR NMM, we study the average behavior of the NMM ensemble. Therefore we generate a sample 3000 time series, one for each NMM. The results are in Figure 8. Here, the leftmost column is the average time series of 1000 Pyr NMM (Figures 8A1, B1, and C1). The middle panel (Figures 8A2, B2, and C2) shows the corresponding power spectrum of the average Pyr NMM outputs. Finally, the rightmost column is the phase plot of the eZJR NMM (Figures 8A3, B3, and C3), each axis reflecting the average of the respective types of neural masses: Pyr, Inh, and Ste.

We got periodic time series and similar spectra, showing all the harmonics independently of the Connectivity Matrix topology (Figures 8A1, A2, B1, B2, C1, and C2). In the phase portrait of our eZJR NMM, the dynamics correspond to stable closed curves but are a bit different among them (Figures 8A3, B3, and C3). As we can see, the FC network dynamics is the simplest and the most similar to a limit cycle(Figure 8C3), while the NN network is the most complex with an extra protuberance (Figure 8A3). The spectra show quite differences among them, which could be associated with the linear spectra used to make the approximation.

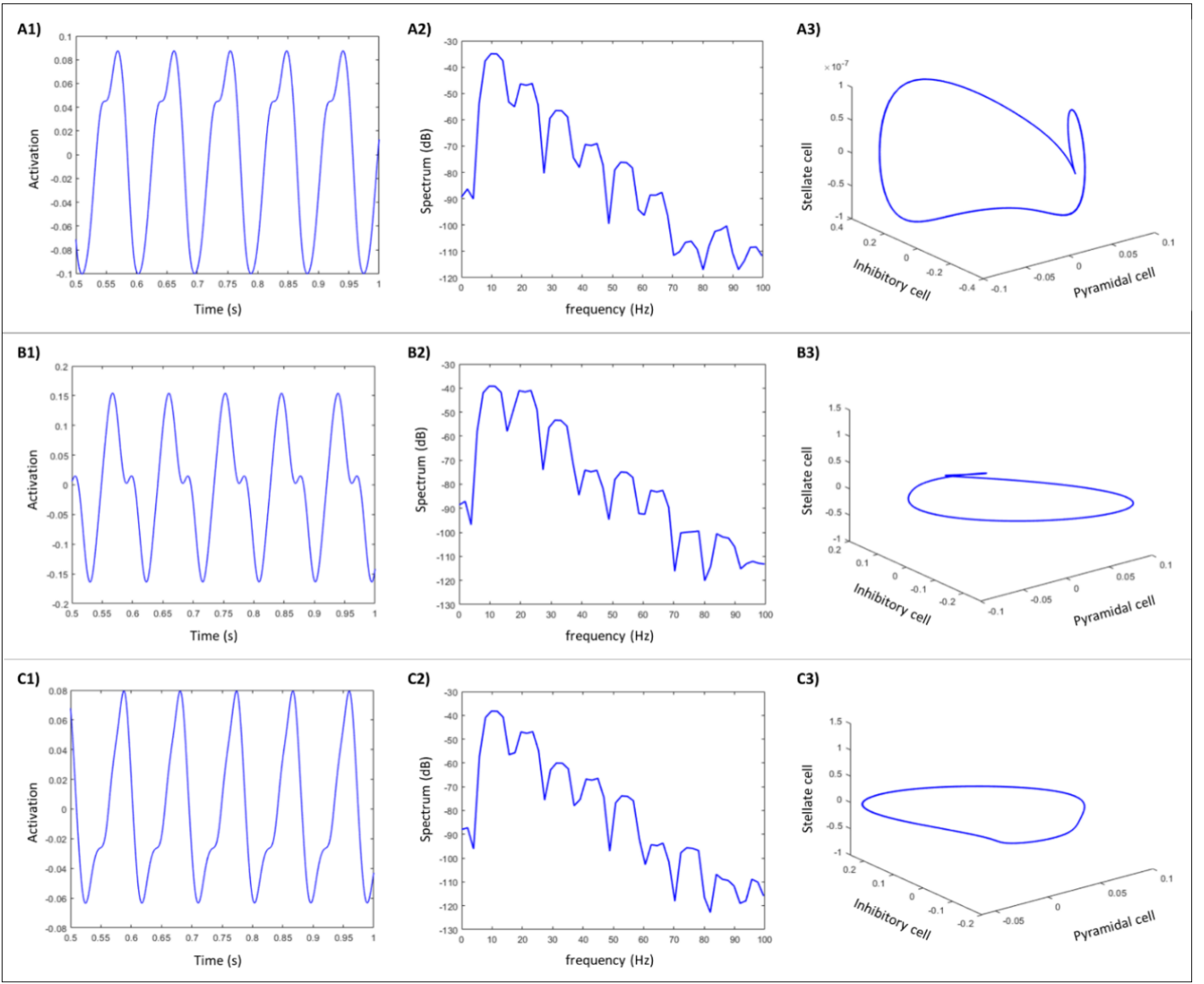

**Figure 8:** Three-panel figure for the connectivity analysis under three types of connections throughout 1000 ZJR cortical columns. Top: correspond to NN network, Middle: represent the SW network and Bottom: describes the FC topology. We make a study for each type of connectivity, based on the generation of the simulated time series (**A1**, **B1** and **C1**), the spectrum (**A2**, **B2**, **C2**) and the oscillatory dynamics (**A3**, **B3**, **C3**).

### 4.2.2 Influence of different types of delay tensors ($D$) in the eZJR NMM dynamics

For the second study at the Population Level, we focus on the influence of time delays into the deterministic ZJR NMM simulation. We introduce three different distributions to represent the activity propagation modeled by a delay tensor $D$. We investigate the influence of the three delay tensors:

- **The instantaneous delay tensor** ($D_1$),
- **The synchronized delay tensor** ($D_2$),
- **The distributed delay tensor** ($D_3$).

We defined them in Section 2.1 and after mentioning them in Section 3. To carry out these simulations, we focused on the SW network.

As we illustrate in Figure 10, the simulated time series still behaves smoothly periodically; perhaps we model a considerable number of unsynchronized ZJR cortical columns (Figure 9A1, B1, and C1). Their spectrums are relatively similar with minor changes, preserving the alpha peak and the harmonics (Figure 9A2, B2, and C2). The trajectories in the phase portrait, although different, are relatively similar, mainly for the first two delay tensors ($D_1$ and $D_2$) employed. They are stable closed curves showing up the stability of the deterministic system.

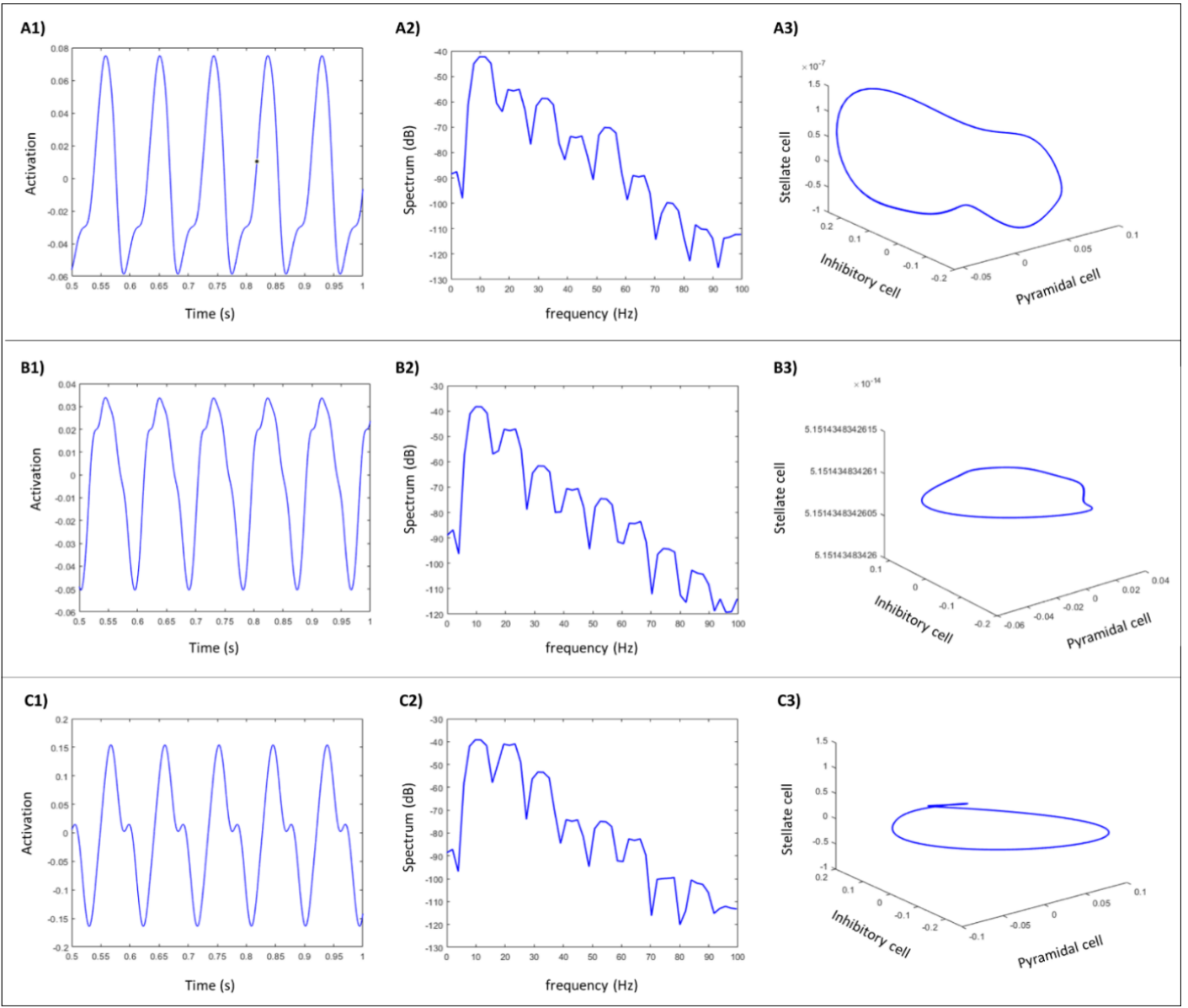

**Figure 9:** Three-panel figure for delay examination throughout 1000 ZJR cortical columns. Top: we simulate the ZJR NMM with instantaneous delay tensor ($D_1$), Middle: we treat time dlays via ynchronized delay tensor ($D_2$) and Bottom: we describe the model using a distributed delay tensor ($D_3$). We make an analysis for each type of delay tensor, based on the generation of the simulated time series (**A1**, **B1** and **C1**), the spectrum (**A2**, **B2** and **C2**) and the oscillatory dynamics (**A3**, **B3** and **C3**).

## 5   Discussion

As mentioned before, despite neural mass models (NMM) being a mainstay of current neuroscience modeling and analysis, current toolboxes suffer from limitations due to the choices of model formulations (9,10,12–15). The most popular methods are based on tightly coupled systems of simultaneous differential equations whose integration quickly becomes computationally prohibitive as each neural mass's complexity or the system's size to simulate increases (17–21). Additionally, usual differential equation techniques that seem popular have two fundamental limitations. i) They use non-optimal methods for integrating stochastic or random systems of differential equations. Some of these methods change the original continuous-time system's dynamical properties (attractors). ii) There is no optimal way to deal with transmission delays between neural masses. There isn't any solution to incorporate into neural mass modeling distributed delays, which is the natural formulation considering the diversity of axonal myelination and diameters.

We leverage a tensor formulation of NMM to solve all these issues:

- Recognizing that neural transmission occurs at a finite speed, we solve the system of equations with a sufficiently small integration step to ensure that input to each neural mass from other masses can be calculated *before* solving the neural mass equation. This approach uncouples the integration of each neural mass, with significant simplifications in the solution of the systems of differential equations (which are now low dimensional) (20,25,26).
- In turn, the integration of each neural mass may now be optimized. We employ the local linearization method, which preserves the dynamical properties of nonlinear systems (27). By decoupling the solution of the equations for each neural mass, we overcome one of the prime difficulties of the LL technique, which involves using the exponential matrix of the Jacobian of the system. Without decupling, this operation becomes computationally demanding at an explosive rate with an increasing number of neural masses. Decoupling the solutions of NMM makes this integration feasible. Simple NMM models such as the Zetterberg-Jansen and Rit (ZJR) (45,57) or Wilson and Cowan (75,76) models can be solved symbolically (before the simulation) to great computational advantage.
- The final ingredient to our formulation accrues from stating the model, not in terms of firing rates but rather in terms of postsynaptic potentials (31). The sigmoid function is applied to the NMM input to produce its output. The input is a linear combination of the past outputs of all other neural masses for a finite time window. It follows readily that the input of the neural mass can be stated as an operation between the Connectome Tensor of delays and the previous output of all other neural masses.
- The Connectome Tensor efficiently encodes any realistic delays between neural masses, including distributed ones, without recourse to ad hoc solutions such as prolonging the postsynaptic potentials.

For single neural ZJR neural mass, the results produced by our toolbox are identical to previous results (45,48,57,77,78) with the same system parameters. The simulation with 1000 cortical columns would have been infeasible with other approaches. The simulations presented in this paper are relatively simple. However, it is striking that different network topologies and delay distributions produce roughly similar results, though essential nuances are evident in the results of the simulations as we tweak the NMM parameters. We plan to use this toolbox to carry out more extensive and realistic modeling, incorporating more biophysical and physiological details.

Two areas need to be studied further. Tensor computations have been optimized for high-performance computer systems. We must modify our code to fully take advantage of the tensor formulation to speed up the computations. This improvement is especially needed when inserting our algorithm into the solution of inverse problems in which simulations must be repeated many times to estimate system parameters.

## 6  Code Availability

The procedures described here are in a toolbox that might enable more realistic large-scale brain modeling, allowing a high-performance simulation of neural dynamic behavior via fast and efficient implementation of the essential core of calculations. Our toolbox and codes are publicly available via Github (https://github.com/CCC-members/Neural_Modelling).

To calculate the time series spectra, we use the Chronux package, an open-source software package for analyzing neural data, available via http://chronux.org/.

## 7  Author Contributions

The contribution of AGM to the presented results was the design of the research, the development of the theory, mathematical demonstration, design of methods, paper preparation and implementation, and the development of the toolbox with DPL and AAG. ML and YW collected the essential references in this field. PVS and MBV senior authors supervised the process and did the final revision. PVS introduced the theoretical background, helped write the article, and guided this work. All authors contributed to the article and approved the submitted version.

## 8 Acknowledgments

The authors would like to thank the support from the National Nature and Science Foundation of China (NSFC) grant No. 61871105 and the CNS program of the UESTC No. Y0301902610100201.